\documentclass[sigconf]{acmart}

\setcopyright{none}
\copyrightyear{2021}
\acmYear{2021}
\acmDOI{}
\acmConference[]{}{}{}
\acmBooktitle{}
\acmPrice{}
\acmISBN{}
\usepackage{amsmath,amssymb,amsfonts}
\usepackage{algorithmic}
\usepackage{graphicx}
\usepackage{textcomp}
\usepackage{xcolor}
\usepackage{url}
\usepackage{mathrsfs}
\usepackage{enumerate}
\usepackage{url}
\usepackage{booktabs} % For formal tables
\usepackage{tabularx}
\usepackage{caption}
\usepackage{subcaption}
\usepackage{balance} 
\usepackage{xspace}
\usepackage[T1]{fontenc}
\usepackage[scaled=0.81]{beramono}
\usepackage{balance}

\usepackage{enumitem}  
\usepackage{listings}
\usepackage{booktabs}
\usepackage{url}
\usepackage{multirow}
\usepackage{array}
\usepackage{color}
\usepackage{float}
\usepackage[misc]{ifsym}
\usepackage{bm}
\usepackage{makecell}
\usepackage{color, colortbl}
\usepackage[font=normalsize]{caption}

\newcommand{\myparagraph}[1]{\textbf{#1.}\quad}

\newcommand{\etal}{{\emph{et al.}}\xspace}
\newcommand{\eg}{{\emph{e.g.}}\xspace}
\newcommand{\ie}{{\emph{i.e.}}\xspace}

\newcommand{\tool}{{Xscope}\xspace}

\begin{document}
	\setcopyright{acmcopyright}
	\copyrightyear{2022}
	\acmYear{2022}
	\acmDOI{XXXXXXX.XXXXXXX}
	
	%% These commands are for a PROCEEDINGS abstract or paper.
	\acmConference[Conference acronym 'XX]{}{}{}
	\acmPrice{15.00}
	\acmISBN{978-1-4503-XXXX-X/18/06}
	
	\title{Xscope: Hunting for Cross-Chain Bridge Attacks}

	\author{Jiashuo Zhang}
	\email{zhangjiashuo@pku.edu.cn}
	\affiliation{%
		\institution{HCST, CS, Peking University}
		\city{Beijing}         % ACM template requires city and
		\country{China}   % country for affiliation
	}
	
	\author{Jianbo Gao}
	\email{gaojianbo@pku.edu.cn}
	\affiliation{%
		\institution{HCST, CS, Peking University}
		\city{Beijing}         % ACM template requires city and
		\country{China}   % country for affiliation
	}

\author{Yue Li}
\email{liyue_cs@pku.edu.cn}
\affiliation{%
	\institution{HCST, CS, Peking University}
	\city{Beijing}
	\country{China}}

\author{Ziming Chen}
\email{chenziming@stu.pku.edu.cn}
\affiliation{%
	\institution{HCST, CS, Peking University}
	\city{Beijing}
	\country{China}
}

\author{Zhi Guan}
	\email{guan@pku.edu.cn}
\affiliation{%
	\institution{Peking University}
	\city{Beijing}
	\country{China}
}
\author{Zhong Chen}
	\email{zhongchen@pku.edu.cn}
\affiliation{%
	\institution{HCST, CS, Peking University}
	\state{Beijing}
	\country{China}
}

	\renewcommand{\shortauthors}{Jiashuo Zhang et al.}
	% keywords, ACM classification and conference information can be omitted for submission
	
	\begin{abstract}
	%Cross-Chain bridges have become the most efficient and popular solution to support asset interoperability between isolated blockchains.   
	%The fragmented ecosystem introduce a significant demand for asset interoperability. 
	Cross-Chain bridges have become the most popular solution to support asset interoperability between heterogeneous blockchains. However, while providing efficient and flexible cross-chain asset transfer, the complex workflow involving both on-chain smart contracts and off-chain programs causes emerging security issues. In the past year, 
	there have been more than ten severe attacks against cross-chain bridges, causing billions of loss. With few studies focusing on the security of cross-chain bridges, the community still lacks the knowledge and tools to mitigate this significant threat. 
	%In the past year, there have been more than ten severe attacks causing billions of loss. 
  To bridge the gap, we conduct the first study on the security of cross-chain bridges.
   We document three new classes of security bugs and 
   %introduce the expected security properties and 
   propose a set of security properties and patterns to characterize them. Based on those patterns, we design \tool, an automatic tool to find security violations in cross-chain bridges and detect real-world attacks. We evaluate \tool on four popular cross-chain bridges. It successfully detects all known attacks and finds suspicious attacks unreported before. A video of \tool is available at https://youtu.be/vMRO\_qOqtXY.
    %Our preliminary evaluation shows  
	%To the best of our knowledge, this is the first study security of cross-chain bridges. 
	\end{abstract}

	
	\maketitle
	\section{Introduction}
	The rise of Bitcoin has spawned thousands of cryptocurrencies and other crypto-assets based on heterogeneous and isolated blockchains. This diverse and fragmented ecosystem
	introduces a significant demand for asset interoperability to allow users to transfer assets between different blockchains.
	%With the efficiency and flexibility gained by introducing off-chain third parties to cross-chain transfer, 
	%By introducing off-chain third parties to coordinate cross-chain tradings, 
	%With the efficiency and flexibility gained from performing with off-chain third-party implementations,
	%Cross-chain bridges have become the most popular method.
	%Cross-Chain bridges have become the most popular solution.
	%and promotes the rapid development of cross-chain bridges. 
	%By introducing an off-chain relayer coordinating different blockchains to provide efficient cross-chain asset transfer, cross-chain bridges
	%With a market cap of over \$300 billion\cite{report}, cross-chain bridges play an increasingly important role in the decentralized ecosystem.
	By introducing off-chain third-party to coordinate cross-chain trading, cross-chain bridges achieve efficient and flexible cross-chain transfers and attract a recent surge.
	%have become the most popular solution with a market cap of over \$300 billion \cite{report}.
	%, they face these difficulties.

While gaining a market cap of billions of dollars~\cite{report},  cross-chain bridges face emerging security issues. Managing assets on different blockchains with complex workflow provides significant advantages to attackers. 
In the past year, more than ten severe attacks leveraging both on-chain and off-chain vulnerabilities caused billions of dollars in damages. 
For example, PolyNetwork~\cite{PolyNetwork} was exploited and lost \$600M due to bugs in on-chain contracts, and THORChain~\cite{THORChain} was attacked three times in 30 days due to bugs in the off-chain relayer.
With these recurrent attacks, the community still lacks knowledge and tools to mitigate the threat.  

	Unfortunately, the security of cross-chain bridges has not been well studied. 
	%Corss-Chain bridges require specific security properties. 
	%A cross-chain bridge attack may 
Cross-Chain bridges involve both on-chain smart contracts and off-chain programs integrating with heterogeneous blockchains, introducing new system architecture and security requirements. However, existing studies \cite{tsankov2018securify,sereum-ndss19,zhang2020txspector,su2021evil} mainly focus on the security of smart contracts on a single chain and target no cross-chain specific security issues. 
	%requires new security properties to guarantee safety. 
	%The recurrent attacks show that the community still lacks understanding of cross-chain attacks.
	%With many reported attacks, the community still lack of 
	%While many previous works have focused on security of smart contract,
	% they all conducted on a single chain and target no cross-chain specific bugs.  
	%With several recent reported attacks, 
%To migrate the threat,
		% received enough attention. 
	%Although several attacks have recently been reported, 
	%the community lack understanding of security properties makes it difficult for developers to survive. 
	%Therefore, reasoning about the correctness of smart contracts before deployment is critical, as is designing a safe smart contract system.
	%has not been well studied. One the one hand, the lack of understanding of cross-chain security makes it difficult for developers to write secure programs.
    Characterizing the security of cross-chain bridges and protecting them against real-work attacks remains an open challenge.	
    %The security properties and 
    %Bug detect in cross-chain bridges remain an open challenge. 
	%This causes the lack understanding of cross-chain bridge
	%Although several 
	%With this significant threats, the community still has little understanding of cross-chain bridge attacks. Although several protocols 
	%Existing 
	%bug detection tools used in practice all focus on smart contracts and decentralized applications on a single chain and target no cross-chain specific attack. 
	
	%have been proposed, 
	%they are limited to a single chain and target no cross-chain specific attacks. 

	% Existing vulnerability detection tools for smart contract 
	%While several bug detection techniques have been proposed for smart contracts and decentralized applications, there are 
	%While several bug detection techniques have been proposed for smart contracts, 
	%received much attention, 
	%To bridge the gap, we formulate the security issues and document new classes of bugs in cross-chain bridges. %With , we propose \tool to detect real-world attacks.
	This paper aims to fill this gap from two perspectives. 
	First, we document new classes of attacks in cross-chain bridges and formulate security properties to enhance the community's understanding.
	Second, we propose \tool to detect security bugs and defend cross-chain bridges against real-world attacks. Specifically, we make the following contributions: 
	%This paper aims to bridge the gap and detect real-world cross-chain bridge attacks. %We characterize the security of cross-chain bridges and propose \tool. 
	%Our contributions are as follows: 
	\vspace{-0.25em}
	\begin{itemize}
	\item  To the best of our knowledge, this is the first study on the security of cross-chain bridges. We present practical challenges causing security issues and document three new classes of security bugs in cross-chain bridges. 
	%We present practical challenges causing these bugs. %We formulate them with logic representations and.
	%We formulate them with a state transition model to characterize practical challenges 
	\item  We introduce a set of security properties and patterns that capture sufficient conditions to characterize cross-chain bridge bugs and detect real-world attacks. 
	Based on them, we propose \tool, an automatic tool providing both run-time monitoring and off-line analyzing functionalities for attack detection and forensics. %cross-chain attacks. 
	\item We preliminarily evaluate \tool on four cross-chain bridges and validate its effectiveness. \tool successfully detects all reported real-world attacks and finds suspicious attacks unreported before. %which demonstrates its effectiveness.
		%We evaluate \tool on four cross-chain bridges related to six severe attacks. The result shows \tool can effectively discover security violations and detect real-world attacks.
	\end{itemize}
\vspace{-0.5em}
	
	\begin{figure*}[t]
		\centering
		\includegraphics[width=1.\textwidth]{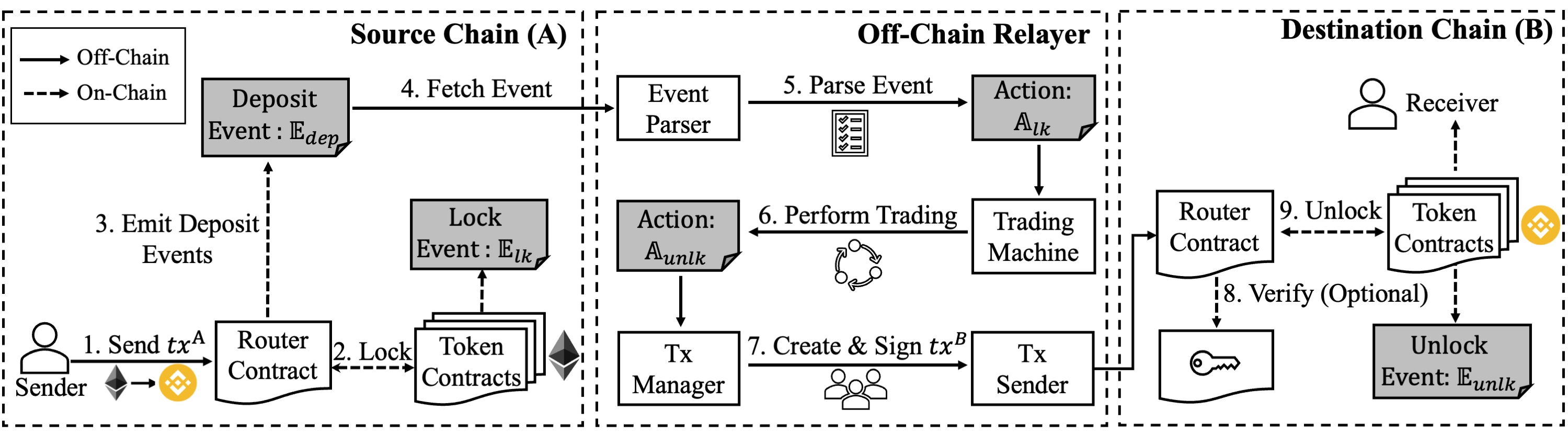} %1.png是图片文件的相对路径
		\vspace{-2em}
		\caption{Cross-Chain Bridge Workflow.  } %caption是图片的标题
		\label{img:bridgel}
				\vspace{-0.5em}
	\end{figure*}
	%\vspace{-5em}
	
	\section{Security Issues in Cross-Chain Bridges}

	\label{sec:bugs}
	The cross-chain bridge is an application acting as an intermediary between users on different blockchains. Users lock various tokens on different blockchains to the bridge, and the bridge takes responsibility for validating these locked tokens, performing cross-chain trading logic like deposit or swap, and unlocking target tokens to users. Correspondingly, a cross-chain bridge consists of two parts, the on-chain router contracts and the off-chain relayer. The router contracts interact with various token contracts and provide on-chain functions, including locking users' tokens and unlocking tokens to users. They also record token transfer information as specific on-chain events to communicate with the off-chain relayer. The off-chain relayer keeps fetching on-chain events from the source chain and coordinates router contracts on the destination chain to finish a cross-chain transfer. 
	%Specifically,  
	%The off-chain relayer coordinates router contracts on different blockchains to finish the cross-chain transfer. 
	%Consequently, the router contracts use events with specific meanings to bridge the big gap between the off-chain relayer and router contracts.
	%The cross-chain bridge is driven by token transfer operations.  
	%Correspondingly, a cross-chain transfer contains three phases: first, the user transfer tokens to router contracts on Chain A. 
	%chain A's router contract locks users' tokens, then the relayer notices the lock action on chain A and authorizes an unlock action on chain B, after that, the router contract on chain B unlocks target tokens to users. 
	%Specifically, the router contracts use on-chain events with specific meanings to communicate with the off-chain relayer. 

	Specifically, we show a detailed workflow of typical cross-chain bridges in Fig~\ref{img:bridgel}. Upon receiving the cross-chain request, the router contract on chain A will call token contracts to lock users' tokens. In this call, the token contract will emit a lock event $\mathit{\mathbb{E}_{lk}}$ and transfer users' tokens to the router contract. 
	After that, the router contract will emit a deposit event $\mathit{\mathbb{E}_{dep}}$ containing detailed information of this lock action such as asset type and amount as proof of locked assets.  
	By parsing $\mathit{\mathbb{E}_{dep}}$, the relayer can learn the lock action $\mathit{\mathbb{A}_{lk}}$ on chain A, authorize the unlock action $\mathit{\mathbb{A}_{unlk}}$ on chain B,
	% according to the trading logic
	and send transactions to the router contract on chain B. 
	%The relayer will f $\mathbb{E}_{dep}$  as proof of locked assets and parse $\mathbb{E}_{dep}$to know the lock action $\mathbb{A}_{lk}$ on chain A and authorize the unlock action $\mathbb{A}_{unlk}$ on chain B. 
	The router contract will verify the transaction sender and other optional proofs like multi-signatures and call target token contracts. Eventually, the token contract emits an unlock event $\mathit{\mathbb{E}_{unlk}}$ and unlocks the target tokens to the users' addresses.
	\vspace{-0.9em}
	\subsection{Bugs in Cross-Chain Bridges}
	Although the lock-then-unlock idea is fairly simple, cross-chain bridges face three practical challenges which have caused severe attacks in reality. First, managing various assets with inconsistent contract interfaces on heterogeneous blockchains introduces bug-prone on-chain logic. Second, the big on-chain-off-chain gap causes complicated off-chain code and leads to potential inconsistency between on-chain router contracts and off-chain relayer. Third, cross-chain bridges' complex and multi-step workflow introduces large attack surfaces, including both traditional and blockchain-specific attacks.
	%Several critical but complex steps become the most vulnerable parts in cross-chain bridges.
	According to these three challenges,  we document and categorize three new classes of security bugs in existing cross-chain bridges\footnote{We summarize existing attacks in cross-chain bridges at https://github.com/Xscope-Tool/Cross-Chain-Attacks} as follows. 
		\begin{table}[t]
		\centering
		\setlength\tabcolsep{1pt}
		\caption{Notations for States in Cross-Chain Bridges. $s$ and $d$ represent the source chain and the destination chain respectively.}
		\vspace{-1em}
		\label{tab:notation}
		\begin{tabular}{@{}lll@{}}
			\toprule
			\textbf{Notation} & \textbf{State} & \textbf{Attributes}               \\ \midrule
			%$\mathit{E}$: Event                 & Chain A  & $\langle \mathit{tx},\mathit{sc}, [\mathit{Attributes}]^+\rangle $                        \\
			$\mathit{\mathbb{E}_{lk}}$               &Lock Event           & $\langle \mathit{tx}^s, \mathit{sc}^s,\mathit{asset}^s,\mathit{amount}^s, \mathit{to}^s\rangle$            \\
			$\mathit{\mathbb{E}_{dep}}$ & Deposit Event          & $\langle\mathit{tx}^s,\mathit{sc}^s,\mathit{asset}^s,\mathit{amount}^s,  \mathit{ID}^d,\mathit{asset}^d,\mathit{to}^d\rangle$ \\
			$\mathit{\mathbb{E}_{unlk}}$& Unlock Event             & $\langle \mathit{tx}^d, \mathit{sc}^d,\mathit{asset}^d,\mathit{amount}^d,\mathit{to}^d\rangle $              \\
			$\mathbb{T}\mathit{(tx)}$               & Trace of $\mathit{tx}$   & $ \{Event|Event.tx=tx\}$                               \\
			$\mathit{\mathbb{A}_{lk}}$ &  Lock Action                & $\langle tx^s,\mathit{ID}^s,  \mathit{ID}^d,\mathit{asset}^s,\mathit{amount}^s,\mathit{asset}^d,\mathit{to}^d\rangle$       \\
			$\mathit{\mathbb{A}_{unlk}}$ & Unlock Action             & $\langle  tx^s, \mathit{ID}^d,\mathit{asset}^d,\mathit{amount}^d,\mathit{to}^d\rangle$       \\
			\bottomrule
		\end{tabular}
			\vspace{-2.2em}
	\end{table}

	\myparagraph{Unrestricted Deposit Emitting (UDE)}
	This bug happens in Steps 2 and 3 at the router contract in Figure~\ref{img:bridgel}. 
	%Typically, the off-chain relayer follows the no lock, no unlock principle and only authorize unlock when see a
	Typically, the router contract should lock senders' tokens before emitting the deposit event $\mathbb{E}_{dep}$. The relayer will regard $\mathbb{E}_{dep}$ as proof of locked tokens and authorize unlocking on the destination chain. 
	%needs to call different token contracts with inconsistent interfaces to lock senders' tokens and then emit deposit events. 
	However, mishandling complex contract interfaces like using unsafe transfer functions may let attackers bypass the lock procedure and trigger a valid deposit event directly. It will provide proof of non-existent locked tokens to the relayer and cause a false deposit on the destination chain. This bug caused Qubit, Meter.io, Wormhole, and Multichain bridges' exploits.
	\begin{table*}[ht]
		\renewcommand\arraystretch{1}
		\centering
		\caption{Security Facts and Inference Rules for in Cross-Chain Bridges. Representations of notations are listed in Table \ref{tab:notation}.   }
		%Informally, $\mathit{V(tx^s)}$ guarantees that if $tx$ contains a valid $\mathit{\mathbb{E}_{dep}}$ from the router contract, it must also lock real asset to the router contract.     
		\vspace{-1em}
		\label{tab:my-table}
		\begin{tabular}{ m{1.9cm} m{5.2cm} m{9.9cm}}
			\toprule
			\textbf{Security Fact}      & \textbf{Intuitive Meaning}   & \textbf{ Inference Rule}                                                                                                                                                                                          \\ \midrule %\mathbb{E}_{lk}
			$\mathit{V(\mathbb{E}_{lk})}$      &Lock Real Asset to Router Contract& $\mathit{\mathbb{E}_{lk}.sc}^s\mathit{ = }\mathit{\mathbb{E}_{lk}.asset}^s \wedge \mathit{\mathbb{E}_{lk}.to}^s \mathit{ = }\mathit{routerContract}^s$                                                                                                        \\
			$\mathit{V(\mathbb{E}_{dep})}$     &Real $\mathbb{E}_{dep}$ Generated by Router Contract& $\mathit{\mathbb{E}_{dep}.sc}^c\mathit{ = }\mathit{routerContract}^s$                                                                                                                                                                       \\
			$\mathit{C(\mathbb{E}_{lk},  \mathbb{E}_{dep})}$      &Same Asset Type and Amount& $\mathbb{E}_{lk}\mathit{.tx}^s\mathit{ = }\mathbb{E}_{dep}\mathit{.tx}^s \wedge \mathbb{E}_{lk}\mathit{.asset}^s\mathit{ =}\mathbb{E}_{dep}\mathit{.asset}^s\wedge\mathbb{E}_{lk}\mathit{.amount}^s\mathit{ = }\mathbb{E}_{dep}\mathit{.amount}^s$                                                        \\
			$\mathit{V(tx^s)}$      &No Lock, No Deposit& $\forall \mathbb{E}_{dep}\in \mathit{\mathbb{T}(tx^s)}, \mathit{V(\mathbb{E}_{dep})} \rightarrow  \exists \mathbb{E}_{lk} \in \mathbb{T}(tx^s),$ $ \mathit{s.t.,V(\mathbb{E}_{lk})}\wedge \mathit{C(\mathbb{E}_{lk},\mathbb{E}_{dep})}$                                                                                                                                         \\
			%\rowcolor{gray!50} $\mathit{RDE}$      & Restricted Deposit Event    & UDE   & $\forall tx^s, VinT(tx^s)   $                    \\ \midrule
			$\mathit{C(\mathbb{A}_{lk},\mathbb{E}_{dep})}$    &Same Asset Type, Amount and Receiver& \makecell[l]{$\mathbb{A}_{lk}\mathit{.asset^s=\mathbb{E}_{dep}.asset^s} \wedge \mathbb{A}_{lk}\mathit{.amount^s=\mathbb{E}_{dep}.amount^s} $\\ $\wedge \mathbb{A}_{lk}\mathit{.asset^d=\mathbb{E}_{dep}.asset^d} \wedge  \mathbb{A}_{lk}\mathit{.to}^d=\mathit{\mathbb{E}_{dep}.to}^d$} \\
			$\mathit{V(\mathbb{A}_{lk})}$   &Consistent $\mathbb{A}_{lk}$ Parsed from Valid tx & $ V(\mathbb{A}_{lk}.tx^s) \wedge\exists \mathit{\mathbb{E}_{dep}} \in \mathbb{T}\mathit{(tx^s)}$ $\mathit{s.t. V(\mathbb{E}_{dep})} \wedge C(\mathbb{A}_{lk},\mathbb{E}_{dep})$                                                                                                                                                                                \\
			$\mathit{C(\mathbb{A}_{lk},\mathbb{A}_{unlk})}$   &Consistent $\mathbb{A}_{unlk}$ based on Valid $\mathbb{A}_{lk}$& $V(\mathbb{A}_{lk})\mathit{\wedge\mathbb{A}_{unlk}.\mathit{ID}^d= \mathbb{A}_{lk}.\mathit{ID}^d\wedge\mathbb{A}_{unlk}.asset^d=}\mathbb{A}_{lk}\mathit{.asset^d\wedge}\mathbb{A}_{unlk}\mathit{.to^d=}\mathbb{A}_{lk}\mathit{.to^d}$                                                                                                                                                                                \\
			%\rowcolor{gray!50}$\mathit{CEP}$      & Consistent Event Parsing  & IEP  &$\forall inA, \exists inTx, s.t. \mathit{VinT(inTx)} \wedge \mathit{VinAT(inA,inTx)}$\\  \midrule
			%$\mathit{CA(inA,outA)}$      & Consistent in and out Actions  & IEP  & \makecell[l]{$\mathit{inA.ID}^d=\mathit{outA.ID}^d \wedge \mathit{inA.asset}^d=\mathit{outA.asset}^d \wedge \mathit{inA.to}^d = \mathit{outA.to}^d$ }  \\  \midrule
			$\mathit{C(\mathbb{A}_{unlk},\mathbb{E}_{unlk})}$&Unlock Right Asset to Right Address& \makecell[l]{$\mathbb{E}_{unlk}\mathit{.asset}^d=\mathbb{E}_{unlk}\mathit{.asset}^d \wedge \mathit{\mathbb{A}_{unlk}.asset}^d=\mathbb{E}_{unlk}\mathit{.sc}^d $\\ $ \wedge \mathit{\mathbb{A}_{unlk}.amount}^d=\mathbb{E}_{unlk}\mathit{.amount}^d \wedge \mathit{\mathbb{A}_{unlk}.to}^d=\mathbb{E}_{unlk}\mathit{.to}^d$} \\ 
			$\mathit{V(\mathbb{E}_{unlk})}$&Authorized Unlock based on Real Lock     &     $\exists \mathbb{A}_{unlk},\mathbb{A}_{lk},s.t., \mathit{C(\mathbb{A}_{unlk},\mathbb{E}_{unlk}) \wedge C(\mathbb{A}_{lk},\mathbb{A}_{unlk})\wedge V(\mathbb{A}_{lk}) } $                                                                                                                                                                                                                                    
			%	\rowcolor{gray!50} $\mathit{AU}$& Authorized Unlocking & UU& $\forall \mathit{UE}, \mathit{AU(UE)}$ 
			%$\mathit{AoutT(outTx)}$& Authorized outTx & UU& $\forall \mathit{UE} \in \mathit{tr(outTx) }, \mathit{AU(UE)}$                                    
			%$\mathit{PoutT(outTx)}$      : outTx with Proof               & $\exists \mathit{A and } \mathit{inTX} \ s.t.\  \mathit{CAinT(A,inTx)} \wedge \mathit{CAoutT(A,outTx)}$                                                                                                                                                                         \\
			%$\mathit{VoutT(outTx)}$      : Valid outTx on Destination Chain & $\sigma$ is a valid signature $ \wedge  \mathit{PoutT(outTx)}$                                                                                                                                                                         
			\\ \bottomrule
		\end{tabular}
	\end{table*}
	\myparagraph{Inconsistent Event Parsing (IEP)}
	%TODO There is a gap
	This bug happens in Step 5 at the event parser in Figure~\ref{img:bridgel} and causes inconsistent off-chain action $\mathit{\mathbb{A}_{lk}}$ and on-chain event $\mathit{\mathbb{E}_{dep}}$. 
	%Due to the on-chain-off-chain gap, the router contract uses deposit event $\mathbb{E}_{dep}$ to communicate with the off-chain relayer which introduces two common bugs. 
	It includes two common mistakes. First, 
	%The reasons are two fold: 
	%In this issue, the parser parses out inconsistent actions with on-chain events. There are two main reasons: 
	the parser may recognize invalid events emitted by malicious contracts as valid deposit events. 
	Second, even with a valid $\mathit{\mathbb{E}_{dep}}$, the parser may parse out invalid lock action $\mathit{\mathbb{A}_{lk}}$ with wrong token types or amount, \eg, recognize a fake token named "ETH" as native ETH. 
	%the valid events but got the wrong asset type or amount.
	%This bug is quite common because a bridge may integrate with many contracts on heterogeneous blockchains, and it is hard for the developer to understand all of them well. For example, 
	This bug resulted in three attacks on THORChain, and one attack on pNetwork.
	%THORChain was exploited three times in 30 days due to this bug. pNetwork's exploit was caused by this bug, too.
	
	\myparagraph{Unauthorized Unlocking (UU)}
	This bug happens in Steps 7 and 8 in Figure~\ref{img:bridgel}. 
	%This bug describes an unlock action on destination chain B 
	%The cross-chain bridge relyes on a logic: no lock, no unlock. 
	%Each unlock event $\mathbb{E}_{unlk}$, must be authorized based on a valid deposit event $\mathbb{E}_{dep}$.
	%Typically, only an authorized party can successfully call the router contract on the destination chain and unlock assets. 
	Typically, only the trusted off-chain relayer can authorize unlock actions on the destination chain.
	However, key leakage caused by traditional cyberattacks or improper access control in on-chain/off-chain codes may allow unauthorized attackers to successfully call the unlock function of the router contract and transfer out assets. This bug caused Robin Bridge, Anyswap, and PolyNetwork's exploits. 
	\vspace{-1em}
	\subsection{Security Properties and Patterns}
	%knowledge: vaultList, routerLis
	% Several critical but complex steps become the most vulnerable parts in cross-chain bridges.
	% and has caused many severe attacks in reality. 
	%In three new classes of attacks,  
	%Those three new classes of attacks  focus on critical but complex steps (Steps 2, 3, 5, 7, 8) in Fig~\ref{img:bridgel}.
	%The unexpected function in critical but complex steps (Steps 2, 3, 5, 7, 8) in Fig~\ref{img:bridgel} 
	%The cross-chain bridge relies on the security properties of each step to guarantee safety and . 
	%which makes several several critical but complex steps become the most vulnerable parts in cross-chain bridges.
	%\eg, the event parser should never regard events generated by non-whitelisted contracts as valid. 
	%Several critical but complex steps become the most vulnerable parts in cross-chain bridges.
	%The cross-chain bridge expected each step work properly to guarantee safety. 
	%The cross-chain bridge relies on the security properties of each step to guarantee safety and . 
	%Cross-Chain bridges has a  relies on the security properties of each step to guarantee safety.
	%Cross-chain bridges require multi-step operations to complete cross-chain transfer and rely on the proper functioning of each step to guarantee security.
	%However, the lack of understanding of security properties in several critical but complex steps has caused severe attacks in reality.
	In this part, we formulate new security issues in cross-chain bridges and propose security properties to detect them.
	
	%First, we need to formalize the sophisticated cross-chain bridge behaviors to 
	We start from an observation that a cross-chain transfer shown in Figure~\ref{img:bridgel} can be characterized as an execution sequence containing a serial of on-chain events and off-chain actions, \ie, $\mathit{\mathbb{E}_{lk}\rightarrow\mathbb{E}_{dep}}$ $\mathit{\rightarrow\mathbb{A}_{lk}\rightarrow\mathbb{A}_{unlk}\rightarrow\mathbb{E}_{unlk}}$.   
	Specifically, $\mathit{\mathbb{E}_{lk}\rightarrow\mathbb{E}_{dep}}$ describes operations on the source chain: first, the token contract transfers users' tokens to the router contract and emits $\mathbb{E}_{lk}$, then, the router contract emit $\mathbb{E}_{dep}$ to aggregate on-chain lock actions into a uniform representation and inform the off-chain relayer. 
	% where 
	%we use $\mathbb{E}_{lk}$ and $\mathbb{E}_{dep}$ to describe router contracts on source chain. 
	%$\mathbb{E}_{lk}$ is emitted by token contracts and shows the lock actions on source chain. $\mathbb{E}_{dep}$ is emitted by the router contract to aggregate on-chain lock actions into a uniform representation and inform the off-chain relayer. 
	$\mathit{\mathbb{E}_{dep}\rightarrow\mathbb{A}_{lk}\rightarrow\mathbb{A}_{unlk}}$ describes the relayers' behaviors, \ie, parse $\mathit{\mathbb{E}_{dep}}$ to learn on-chain actions $\mathit{\mathbb{A}_{lk}}$ and authorize unlock action  $\mathit{\mathbb{A}_{unlk}}$ on the destination chain. $\mathit{\mathbb{A}_{unlk}\rightarrow\mathbb{E}_{unlk}}$ shows operations on destination chain, \ie,
	receive relayer's transactions and unlock target tokens to users. 
	Furthermore, we transform states in the sequence into representations in Table~\ref{tab:notation}. For better compatibility, the representations follow the minimum principle and only contain essential attributes that can be easily derivated from most cross-chain bridges. 
	
	After simplifying the complex workflow into a four-step sequence from $\mathit{\mathbb{E}_{lk}}$ to $\mathit{\mathbb{E}_{unlk}}$, we formulate the security issues with logical constraints in the execution sequence.
	%and makes formalization and furture analysis feasible. 
	%Then, we formulate security violations in an execution sequence with logic representations. 
	Specifically, we introduce two types of security facts in Table~\ref{tab:my-table} to describe the constraints, \ie, Validity facts and Consistence facts, denoted by $\mathit{V}$ and $\mathit{C}$ respectively.
	%To describe the expected behavior of cross-chain bridges, 
	%we conduct two types of security facts, \ie, Validity ($\mathit{V}$) facts and Consistence ($\mathit{C}$) facts, and list them in Table~\ref{tab:my-table}. 
	Validity facts define whether a state is valid to be part of the execution sequence. 
	For example, $\mathit{V(\mathbb{E}_{dep})}$  defines that only $\mathit{\mathbb{E}_{dep}}$ emitted by the router contract can occur in an execution sequence and finally cause an unlock on the destination chain while an invalid $\mathit{\mathbb{E}_{dep}}$ emitted by malicious contracts should never take part in a normal sequence.
	%will lead to a security violation. 
	% and a sequence contains an invalid $\mathbb{E}_{dep}$ represents that the off-chain relayer may falsely parse $\mathbb{E}_{dep}$ emitted by malicious contracts and lead to a security violation. 
	%Hence, if an execution sequence contains 
	%a $\mathbb{E}_{dep}$ generated 
	Consistence facts define the expected relations between two states. For example, $\mathit{C(\mathbb{E}_{dep},\mathbb{A}_{lk})}$ requires exactly the same asset type and amount in $\mathit{\mathbb{E}_{dep}}$ and $\mathit{\mathbb{A}_{lk}}$, \ie, the off-chain relayer correctly learns what asset is locked to the router contract. 
	%Violating this fact demonstrates 
	%there is no inconsistency between off-chain relayer and on-chain router contracts.
	
	Based on the security facts we conduct, 
	%we can detect security violations in any execution sequences of cross-chain bridges and find potential security bugs.  
	%and detect security violations.
	%Specifically, in Table~\ref{tab:property}, 
	we introduce three security properties and patterns in Table~\ref{tab:property} to detect three new classes of cross-chain bridge bugs. Respectively, 
	$\mathcal{RD}$ ensures every valid deposit event has a corresponding valid lock event to detect the deposit-without-lock violations in the UDE bug.
	% characterizes the UDE bug by ensuring every valid deposit event has a corresponding valid lock event so the attacker can never bypass the lock procedure and emit deposit event directly. 
	$\mathcal{CP}$ detects the IEP bug
	by ensuring every off-chain action  $\mathit{\mathbb{A}_{lk}}$ is parsed from a valid and consistent $\mathit{\mathbb{E}_{dep}}$. %to characterizes the IEP bug.
	And $\mathcal{AU}$ 
	%characterizes the UU bug by 
	ensures every unlock event must be caused by an authorized unlock action $\mathit{\mathbb{A}_{lk}}$  from the off-chain relayer to characterize the UU bug. 
	%These three properties capture sufficient conditions
	%what consistence requiments 
	%whether two states follow an expected consistence requirements. s
	% cross-chain transfer sequence.
	%We shown the on-chain events and off 
	%generated by the token contract to 
	%In this part, 
	%we propose security properties to characterize expected behaviors of these steps and 
	%these three classes of attacks and 
	%conduct several patterns to find security violations and detect UDE, IEP, UU attacks mentioned above. 
	%we focus on several critical but complex steps including Steps 2, 3, 5, 7 and 8 in Fig~\ref{img:bridgel} have become the most vulnerable part in cross-chain bridges. 
	
	%In table~\ref{tab:notation}, we show the logic representations. 

	%so that it can be compatible to different bridges with little effort. 
	%For example, the lock/unlock event can be derived from transfer events in the most polular ERC-20/BEP-20 specifications.

	\begin{table}[]
		\centering
		\renewcommand\arraystretch{1}
		\setlength\tabcolsep{2.pt}
		\caption{Security Properties and Patterns to Detects Attacks}
				\vspace{-1em}
		\label{tab:property}
		\begin{tabular}{@{}lll@{}}
			\toprule
			\multicolumn{1}{c}{\textbf{\begin{tabular}[c]{@{}c@{}}Prop.\end{tabular}}} & \multicolumn{1}{c}{\textbf{Description}} & \multicolumn{1}{c}{\textbf{\begin{tabular}[c]{@{}c@{}} Pattern\end{tabular}}} \\ \midrule
			$\mathcal{RD}$&  Restricted Deposit: No Lock, No Deposit  &        $\forall tx^s, V(tx^s) $          \\ 
			$\mathcal{CP}$	&  Consistent Parsing: Get Real Actions  &          $\forall \mathbb{A}_{lk}, \mathit{V(\mathbb{A}_{lk})}$        \\
			$\mathcal{AU}$	&  Authorized Unlock: Unlock as Expected  &            $\forall \mathbb{E}_{unlk}, \mathit{V(\mathbb{E}_{unlk})}$        \\ \bottomrule
		\end{tabular}
	\vspace{-2em}
	\end{table}
	\vspace{-0.5em}
	\section{Tool Overview}
	%In this section, we first formalize cross-chain bridges with on-chain events and off-chain actions and transform them into logical representations.   
	
	%Managing various assets on heterogeneous blockchains with on-chain and off-chain codes written in different programming languages introduces new attack surfaces and has caused many severe attacks. 
	%Cross-Chain bridges work in a multi-phase and log-driven way. The router contract 
	Based on these security properties and patterns, we propose \tool, an automatic tool to detect cross-chain bridge attacks. 
	%to find security violations in cross-chain bridges and detect real-world attacks. 
	\tool provides two modes to use: runtime monitoring and offline analyzing. 
	In runtime monitoring mode, it is deployed as an extension of the off-chain relayer to enforce security properties and abort malicious requests. Specifically, the off-chain relayer will additionally call \tool after generating outgoing transactions at step 7 in Figure~\ref{img:bridgel}. 
	%Before sending outgoing transactions, the relayer will call \tool to verify them.  
	Then, \tool will pre-execute the outgoing transactions to get complete execution sequences from $\mathit{\mathbb{E}_{lk}}$ to $\mathit{\mathbb{E}_{unlk}}$ and detect security violations in them. If a violation is detected, the relayer will abort the outgoing transactions to prevent potential loss.
	%In runtime monitoring mode, it is deployed on the off-chain relayer and fetch the current exec
	%keeps checking the execution sequence of incoming cross-chain requests, pre-executes the outgoing unlock transactions and rejects them when security violations are detected.
	In offline analyzing mode, \tool passively analyses historical requests and issues warnings for suspicious execution sequences.
	
	Correspondingly, \tool provides twofold benefits. With runtime monitoring mode, it protects existing bridges from real-world attacks with updatable and flexible security patterns.
	%First, it supports runtime monitoring with negligible overhead to detect new classes of attacks. It provides updatable and flexible configuration and can be extended to support other (even non-security) requirements by customizing user-specific patterns.
	With off-line analyzing mode, it provides efficient forensics and accurate error location for bridges to investigate past or ongoing attacks.
	% more timely and take countermeasures like freezing attackers' accounts
	%make patches like freezing the attackers' accounts 
	%as soon as possible.
	
	%\tool supports updatable patterns.  It is also flexible and can be extended to support other (even non-security related) analysis requirements by customizing user-specific patterns
	
\begin{figure}[]
		\centering
		\includegraphics[width=.4\textwidth]{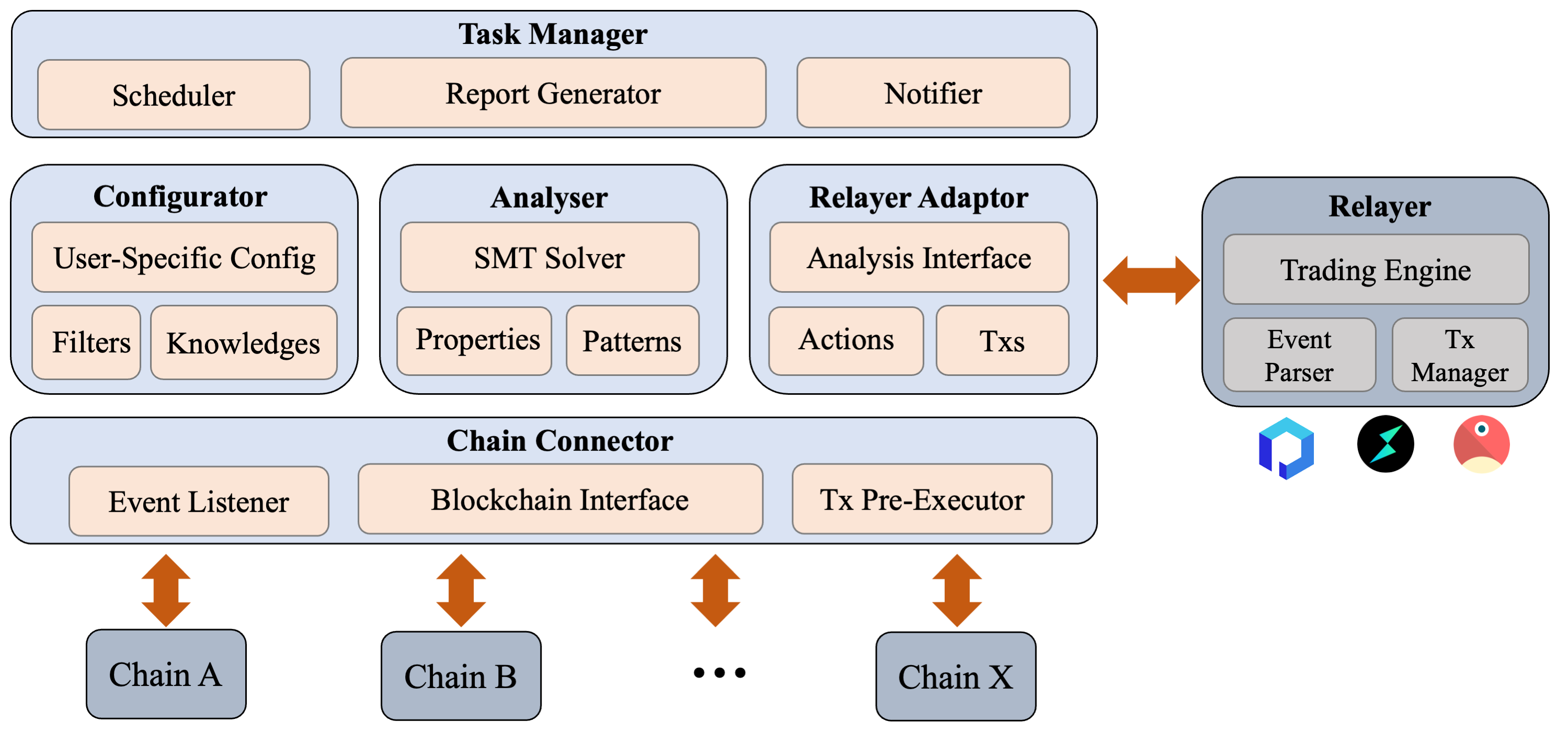} %1.png是图片文件的相对路径
		\vspace{-0.8em}
		\caption{Architecture of \tool} %caption是图片的标题
		\vspace{-1.3em}
		\label{img:arch}
	\end{figure}

	Figure~\ref{img:arch} shows the architecture of \tool. 
	%As shown in Fig~\ref{img:arch}, \tool consists of five modules. 
	The Chain Connector and Relayer Adaptor provide a consistent interface above different blockchains and relayers. They transform on-chain events and off-chain actions into logical representations and feed them into the Analyzer. 
	% interacts with the relayer and transforms the relayer's execution data into logic representations for analysis.  
	%The relayer handles in-transactions on one chain and generates out-transactions on another. 
	%In runtime mode, before submitting out-transactions to the destination chain, the relayer will pre-execute the transaction and get events on both chains and off-chain actions to the Analyzer.
	The Analyzer uses Z3 as SMT Solver to detect violations in each execution sequence.
	% checks each execution sequence and detects violations of security properties using Z3 as SMT Solver. 
	%Furthermore, 
	Additionally, the Configurator is introduced to support user-specific configurations, \eg, filtering out transactions from specific addresses or importing external knowledge like a blacklist. The Task Manager is designed to schedule analysis/monitor tasks, generate reports, and send notifications to users. 
	%and provides on-chain events or transaction pre-executor depend on whether the transaction is. %In off-line mode, all transactions are already finished 
	%In detail, it feteches on-chain events from chain adaptor and off-chain actions from relayers. In the off-line mode, the  while in runtime mode, the analyser need to pre-execute locally to fetch on-chain events. 
	%In off-line mode, the analyser feteches on-chain events and off-chain actions and detect violations of security properties.
	%In run-time node, as the router will call analyser to verify out transactions before submitting them to destination chain, the analyser needs to pre-execute the out transactions to detect security violations based on the results. 

\begin{figure}[tp]
		\centering
		\includegraphics[width=.46\textwidth]{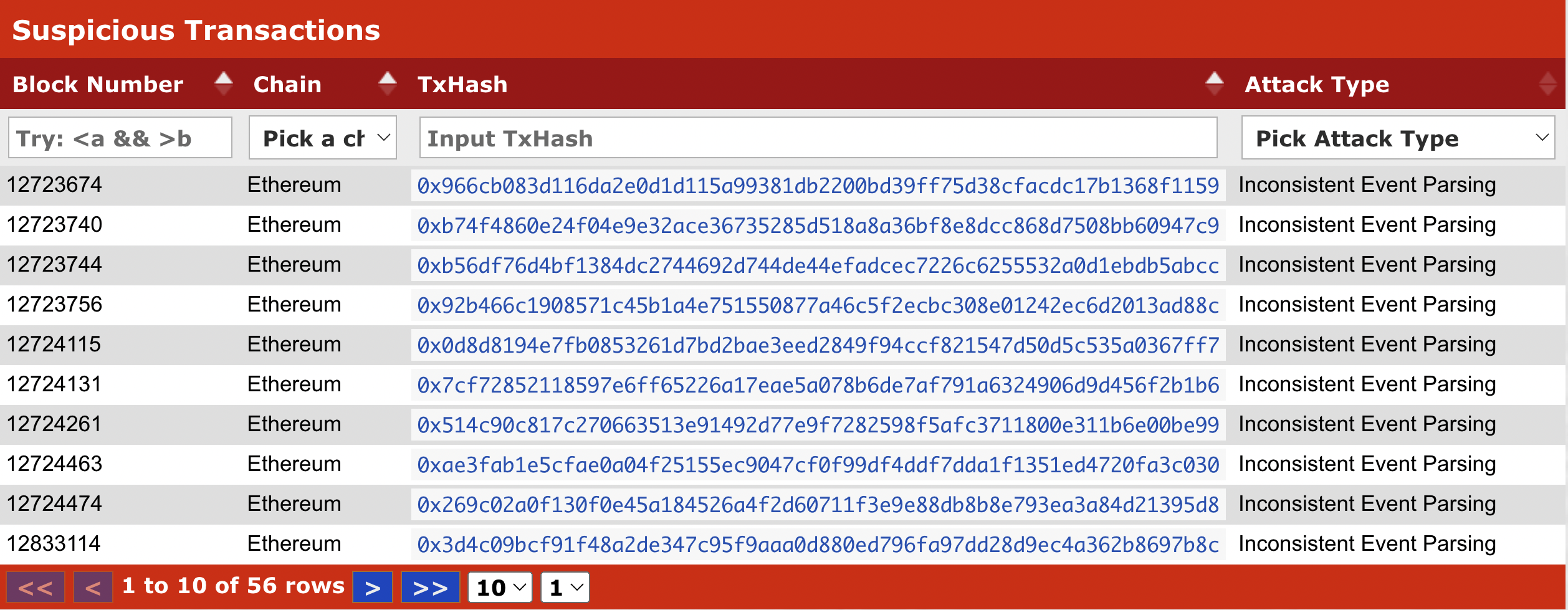} %1.png是图片文件的相对路径
		\vspace{-0.5em}
		\caption{The Output Interface of \tool} %caption是图片的标题
		\vspace{-1.2em}
		\label{img:output}
	\end{figure}

	An example output of \tool is shown in Figure~\ref{img:output}. In this case, we collect over 2 million historical transactions from a cross-chain bridge named THORChain\footnote{The data is available at https://github.com/Xscope-Tool/Data} and use \tool to detect security violations in them. 
	%all historical transactions (about 20k) before April 1, 2022. of 
	\tool shows a list containing 56 suspicious transactions clustered in three time periods (Block 12723674-12724474, 12833114-12833448, and 12878663-12878671 on Ethereum), each labeled with relevant bugs. \tool also supports searching, filtering, and sorting functions based on attributes like block number and transaction hash, which can help cross-chain bridges investigate those transactions efficiently. 
	\vspace{-0.5em}
	\section{Preliminary Evaluation}
	In this section, we conduct a preliminary evaluation of \tool to validate its effectiveness in finding cross-chain bridge attacks. Specifically, we use \tool to analyze four popular cross-chain bridges between Ethereum and Binance Smart Chain related to six known cross-chain bridge attacks.
	\vspace{-1em}
	\begin{table}[h]
		\renewcommand\arraystretch{1}
		\setlength\tabcolsep{1.25pt}
		\caption{Evaluation Results of \tool}
				\vspace{-1em}
		\label{tab:results}
		\begin{tabular}{lccrccc}
			\hline
			\multicolumn{1}{c}{}                         &                                                                                                                                                 &                                                                           & \multicolumn{3}{c}{Transactions Detected}                                           \\ \cline{4-6} 
			\multicolumn{1}{c}{\multirow{-2}{*}{Attack}} & \multirow{-2}{*}{\begin{tabular}[c]{@{}c@{}}Bug  Type\end{tabular}}  & \multirow{-2}{*}{\begin{tabular}[c]{@{}c@{}}Transactions \\ Reported\end{tabular}} & {Reported}  & { Unreported} & All \\ \hline
			THORChain \#1                                & IEP                                                                              & 6                                                                         & {\color[HTML]{036400} 6 (100\%)}  & \textbf{\color[HTML]{00009B} +3}          & 9   \\
			THORChain \#2                                & IEP                                                              & 41                                                                       & {\color[HTML]{036400} 41  (100\%)} & \textbf{\color[HTML]{00009B} -}          & 41  \\
			THORChain  \#3                               & IEP                                                                               & 6                                                                         & {\color[HTML]{036400} 6 (100\%)}  & \textbf{\color[HTML]{00009B} -}          & 6   \\
			pNetwork                                     & IEP                                    &3                                                  & {\color[HTML]{036400} 3 (100\%)}  & \textbf{\color[HTML]{00009B} -}          & 3   \\
			Anyswap & UU & 4 &{\color[HTML]{036400}4 (100\%)} &\textbf{\color[HTML]{00009B} -}   & 4\\
			Qubit  Bridge                                      & UDE                                                                              & 16                                                                        & {\color[HTML]{036400} 16 (100\%)} & \textbf{\color[HTML]{00009B} +4}          & 20  \\ \hline
		\end{tabular}
	\vspace{-1em}
	\end{table}

	We compare detected suspicious transactions given by \tool with the officially reported attack transactions\footnote{The official reports of these attacks and detailed comparison are available at https://github.com/Xscope-Tool/Results}.   
	The results in Table~\ref{tab:results} show that \tool not only successively detects all reported attack transactions covering three new classes of attacks in cross-chain bridges, but also detects several unreported attack transactions. 
	%It demonstrates the effectiveness 
	In particular,  \tool detects an unreported suspicious transaction 43 days before the major attack in Qubit Bridge on Jan 27, 2022. If this suspicious and unnoticed transaction had been detected just in time, the major attack causing 80M dollars loss might have been prevented. We have reported our findings to the Qubit team. 
	%, we could have prevented subsequent attacks and avoided millions of dollars in losses.
	%It demonstrates the attack was 
	% in Qubit Bridge
	%and has a. In particular,  
	
	% WAT, WAA, IE, IE
	\section{Related Work}
	%Smart Contract Security have been well-studied in recent years. 
	Attack detection for smart contracts \cite{tsankov2018securify,sereum-ndss19,zhang2020txspector} and decentralized applications (DApp) \cite{su2021evil} have been widely discussed over the past years. Rodler \etal \cite{sereum-ndss19} introduce Sereum to protect contracts against re-entrancy attacks based on runtime monitoring. 
	%Chen \etal \cite{soda} propose an online detection framwork for smart contracts. 
	Zhang \etal \cite{zhang2020txspector} propose a tool to detect Ethereum attacks using user-specific detection rules. 
	%Zhang \etal \cite{zhang2021dharcher} propose a test framework to detect two types of bugs in DApp. 
	Su \etal \cite{su2021evil} propose DEFIER to automatically investigate on-chain attack incidents. 
	To our knowledge, they all focus on a single chain and target no cross-chain specific attack. 
	In addition, several previous studies focus on the security of atomic cross-chain swap protocols \cite{herlihy2018atomic,tsabary2021mad} or generic cross-chain architectures like Polkadot \cite{wood2016polkadot} and Cosmos \cite{cosmos},
	%In addition, several studies \cite{tsabary2021mad,xue2021hedging,xu2021game} focus on the security of cross-chain atomic swap protocols \cite{herlihy2018atomic}, 
	which are orthogonal to application-level cross-chain bridges we discuss.  
	%Unlike our work, they focus on a single chain and 
	%They focus on 
	%However, then mainly focus on smart contracts on a single chain and cannot be extended to cross-chain bridges. 
%	\myparagraph{Cross-Chain Security}
	%While the smart contract security has been well studied,  Cross-Chain bridge is a relatively recent areas with fewer works.
	%(TODO)The results demonstrate that \tool not only successively detects all reported attack transactions covering three new classes of attacks in cross-chain bridges.
	%	but also detects several unreported attack transactions. In particular,  \tool detects a suspicious transaction 43 days before the major attack in Qubit on Jan 27, 2022, which means that there may be an unnoticed and small attack before the major attack in Qubit and if it was detected in time, the major attack causing 80M dollars loss could have been prevented. We have reported our findings to the Qubit team for further investigation. 
	\section{Conclusion}
	In this paper, we conduct the first study on the security of cross-chain bridges. We document three new classes of security bugs in cross-chain bridges and introduce a set of security properties and patterns to characterize them. We propose \tool as an automatic tool to find security violations and detect real-work attacks in cross-chain bridges. Our preliminary evaluation demonstrates the effectiveness of \tool. 
%	\item  To the best of our knowledge, we first document and categorize security bugs in cross-chain bridges. We formulate the problem with a state transition model and present practical challenges causing these bugs. 
%\item  We introduce a set of security properties and patterns that capture sufficient conditions to characterize cross-chain bridge bugs and detect real-world attacks. 
%Based on them, we propose \tool, an automatic tool providing both run-time monitoring and passive analysis functions for attack detection and forensics. %cross-chain attacks. 
%\item We preliminarily evaluate \tool on four cross-chain bridges. \tool successfully detects all reported real-world attacks and finds potential attacks unreported before which demonstrates its effectiveness.

\vspace{-0.5em}

	\section*{Acknowledgement}
	This work was supported by National Key Research and Development Program of China (2020YFB1005802), National Natural Science Foundation of China (62172010), and Beijing Natural Science Foundation (M21040).

\vspace{-0.5em}
	
	% ----------------------------------------------------------------
	% you can include the acknowledgements in the source, but `anonymous' option will hide them
	
	% ----------------------------------------------------------------
	% use ACM-Reference-Format for the references
	\bibliography{xscope}
	\bibliographystyle{ACM-Reference-Format}
	
	%\begin{appendix}
	
	%\end{appendix}
	
\end{document}